\documentclass[a4paper,fleqn,usenatbib, letters, onecolumn, oneuseAMS]{mnras}
\usepackage[T1]{fontenc}
\usepackage{ae,aecompl}
\usepackage{rotating}
\usepackage{caption}
\usepackage{longtable}

\usepackage{graphicx}	
\usepackage{amsmath}	
\usepackage{amssymb}	
\usepackage{color}

\title[]{Spectral and timing analysis of the bursting pulsar GRO J1744-28 with {\it RXTE} observations}

\author[L., Ji et al.]{%
	L. Ji$^{1}$\thanks{E-mail: ji.long@astro.uni-tuebingen.de},
	A. Santangelo$^{1}$\thanks{E-mail: andrea.santangelo@uni-tuebingen.de},
	S. Zhang$^{2}$\thanks{E-mail: szhang@ihep.ac.cn},
	V. Doroshenko $^{1}$,
	V. Suleimanov $^{1, 3}$,
	\newauthor 
	L. Ducci  $^{1,4}$,
	P. Kretschmar$^{5}$,
	R. Doroshenko $^{1}$\\
	$^{1}$ Institut f\"ur Astronomie und Astrophysik, Kepler Center for Astro and Particle Physics, Eberhard Karls Universit\"at, Sand 1, 72076 \\
	T\"ubingen, Germany\\
	$^{2}$Key Laboratory for Particle Astrophysics, Institute of High Energy Physics, Beijing 100049, China\\
	$^{3}$Kazan (Volga region) Federal University, Kremlevskaya str. 18, 42008 Kazan, Russia,\\
	$^{4}$ ISDC Data Center for Astrophysics, Universit\'e de Gen\`eve, 16 chemin 	d'\'Ecogia, 1290 Versoix, Switzerland\\
	$^{5}$European Space Astronomy Centre (ESA/ESAC), Science Operations Department, Villanueva de la Ca\~nada (Madrid), Spain \\
}

\date{Accepted XXX. Received YYY; in original form ZZZ}

\pubyear{2016}

\begin{document}
\label{firstpage}
\pagerange{\pageref{firstpage}--\pageref{lastpage}}
\maketitle

\begin{abstract}
We analyzed {\it RXTE/PCA} observations of the bursting pulsar GRO J1744-28 during its two outbursts in 1995-1997.
We have found a significant transition, i.e., a sharp change, of both spectral and timing properties around a flux of $\rm 10^{-8}\ ergs\ cm^{-2}\ s^{-1}$  (3-30\,keV),  which corresponds to a
 luminosity of 2--8$\rm \times 10^{37}\ ergs\ s^{-1}$ for a distance of 4--8\,kpc.
We define the faint (bright) state when the flux is smaller (larger) than this threshold.
In the faint state, the spectral hardness increases significantly with the increasing flux, while remaining almost constant in the bright state.
In addition, we find that the pulsed fraction is positively related to both the flux and the energy (<30\,keV) in both states.
Thanks to the very stable pulse profile shape in all energy bands, a hard X-ray lag could be measured.
This lag is only significant in the faint state (reaching $\sim$ 20\,ms between 3-4\,keV and 16-20\,keV), and much smaller in the bright state ($\lesssim$\,2ms).
We speculate that the discovered transition might be caused by changes of the accretion structure on the surface of the neutron star, and the hard X-ray lags are due to a geometry effect.

\end{abstract}



\section{Introduction}
GRO J1744-28 is a transient low mass X-ray binary (LMXB) system discovered near the Galactic center by \citet{Fishman1995}.
Three main outbursts have been reported since its discovery, i.e., from the end of 1995 to the beginning of 1996,  from the end of 1996 to April 1997, and from January to June 2014.
The X-ray luminosity during the outbursts reaches ${\rm {10}^{37}-{10}^{38}\, ergs\ s^{-1}}$ \citep{Nishiuchi1999, Degenaar2014, DAI2015, Younes2015}, which is several orders of magnitude brighter than that in the quiescent state (${\rm {10}^{33}-{10}^{34}\, ergs\ s^{-1}}$) \citep{Wijnands2002, Daigne2002}.
Plenty of X-ray bursts have been reported for GRO J1744-28, which have been attributed to spasmodic accretion, making the source the second object showing so-called type-II bursts besides the Rapid Burster \citep{Kouveliotou1996}.
A truncated accretion disc has been reported \citep{Degenaar2014, DAI2015}, which might be related to these bursts \citep{Eijnden2017}.
The magnetic field of GRO J1744-28 is thought to be $\sim$ ${10}^{11}$-${10}^{12}$\,G as it follows from the detection of a cyclotron resonance scattering feature (CRSF) between 4 and 5\,keV \citep{Doroshenko2015, DAI2015}.
A similar consistent value has been as well obtained considering the interaction between the accretion disc and the magnetosphere  \citep{Cui1997, Degenaar2014, Younes2015}.

The X-ray spectrum of the source has been investigated using observations of {\it ASCA}, {\it Chandra}, {\it XMM-Newton}, {\it NuSTAR}, and {\it BeppoSAX}  \citep{Nishiuchi1999, Degenaar2014, Younes2015, DAI2015, Doroshenko2015}.
Generally, the spectrum can be fitted with a phenomenological model, consisting of a cutoff power-law and sometimes an additional blackbody component with a temperature of < 1\,keV.
A broad emission line around 6-7\,keV has been reported.
This has been suggested to be a reflection component due to iron lines reprocessed in the accretion disc \citep{Nishiuchi1999, Degenaar2014,DAI2015}.
Soon after the discovery of the source, coherent pulsations, with a frequency of $\sim$ 2.1\,Hz, were found \citep{Finger1996}.
Pulsations show almost a perfect sinusoidal modulation, with a small harmonic content harmonics \citep[see e.g., ][]{Giles1996, DAI2016}.
Generally, the pulsed fraction increases with energy, except for the energy range of 6-7\,keV, due to the contribution of the non-pulsed iron line \citep{DAI2015, Doroshenko2015}.
\citet{DAI2016} reported an energy-dependent phase lag of the hard X-rays (2.3-10\,keV) with respect to the soft 0.3-2.3\,keV energy band.
The hard X-ray lag increases with the energy, and reaches 12\,ms between 6-6.4\,keV, and then drops to 7\,ms followed by a plateau of $\sim$ 9\,ms when the energy is larger than 8\,keV \citep[for details, see Figure 2 in][]{DAI2016}.
The authors have explained the lags by using a Compton reverberation mechanism, i.e., they have suggested the existence of a Compton cloud with a size of $\sim$ 120\,$R_{\rm g}$, compatible with the magnetospheric radius, in which hard X-rays would scatter before escaping.
{Extensive observations of the source revealed, therefore, rich phenomenology, however, the relationship between the spectral and timing properties of GRO J1744-28 is still largely unexplored, especially regarding to their co-evolution along the outbursts.
This information is helpful for constraining the radiation process in accreting pulsars and could shed light on properties of the emission region and their evolution with luminosity.
In this paper, we attempt to close this gap and systematically study the spectral and timing properties of the source over a wide luminosity range during the outbursts.
}

\section{Observations and data analysis}
\begin{figure}
 	\centering
 	\includegraphics[width=4.5in]{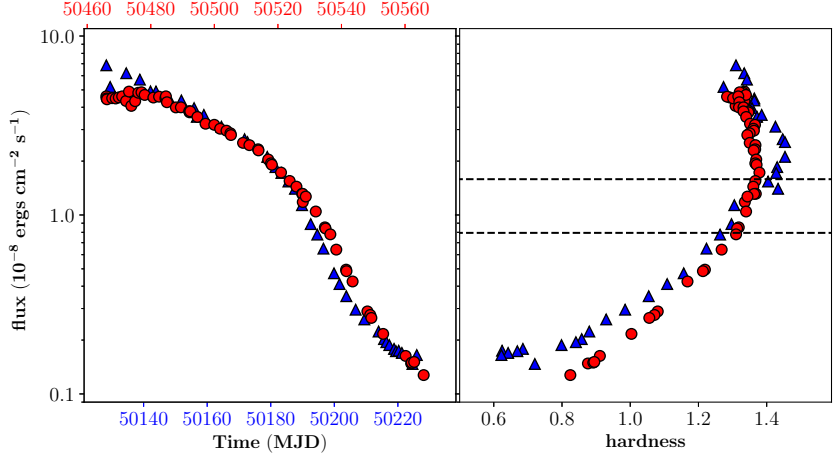}
 	\caption{Long-term lightcurve (left) and hardness-intensity diagram (right) of GRO J1744-28. We show the data of the first and second outbursts in blue and red, respectively. In the left panel, their time scales are the same but with an offset of 337.75\,days. The energy range of the flux is 3--30\,keV. The hardness is defined as the count rate ratio of 10--20\,keV and 4--10\,keV.
 	The dashed lines represent the flux equals to 0.8 and 1.6 $\times 10^{-8}$ $\rm ergs\ cm^{-2}\ s^{-1}$, respectively,  corresponding to the transitional flux in the timing analysis.
 	}
 	\label{lightcurve_and_HID}
 \end{figure}

To achieve this goal, we use the \textit{RXTE/PCA} observations of GRO J1744-28 during its first two outbursts in 1995-1997 (see Figure~\ref{lightcurve_and_HID}).
Many X-ray bursts were identified \citep[see e.g.,][]{Kouveliotou1996} during this period. However, here we focus on the analysis of persistent emission properties, so the bursts were filtered out in the following analysis.
To avoid the influence of bursts on the persistent emissions \citep[e.g.,][]{Stark1996}, we ignored data between 100\,s prior to a burst peak and 500\,s after the burst peak.
In the timing analysis, a data mode with a high time resolution (compared to its spin 2.1\,Hz) was required.
Therefore, only observations, having Event ("E\_125\_$\mu$s\_64") or Generic Binned ("B\_16ms\_64M") data mode, were used.
Totally, \textcolor{black}{100} observational IDs were selected, with an averaged exposure time of $\sim$ 3000\,s.
During the analysis, we used \textit{Heasoft v.6.19} to extract the lightcurves and spectra and to estimate the corresponding background.
The spectral analysis was performed by using \textit{Xspec v.12.9.0}.
For the spectral analysis, the energy band of 3-30\,keV was used.
A systematic error of 1\% was added.
All uncertainties quoted are at a 90\% confidence level, unless indicated otherwise.

\section{Results}
The long-term evolution of outbursts in GRO J1744-28 is shown in Figure~\ref{lightcurve_and_HID}, where the flux is estimated by fitting the spectrum of each observational ID in the energy range of 3-30\,keV (see below).
We overplotted the lightcurves of the two outbursts for the sake of comparison, and found that they were similar.
There is another short outburst around MJD 50280 having a much weaker peak flux.
However, considering the distinct behaviour, we can not exclude that this weak outburst might originate from other sources located in the field of view of \textit{PCA}, close to the GRO J1744-28.
So we exclude these data in the following.
First we investigated the hardness-intensity diagram as a robust probe of the spectral evolution along the outbursts, where the hardness is defined as the (background-subtracted) count rate ratio of 10-20\,keV to 4-10\,keV.
We note that since our sample spans different gain epochs of \textit{PCA},
a slightly different energy-to-channel conversion\footnote{For details, please see https://heasarc.gsfc.nasa.gov/docs/xte/e-c\_table.html} has to be considered.
In practice, we used the units of Crab, i.e., normalized the lightcurves by using the nearest Crab observation in the same energy band and gain epoch.
This method is widely used to extract lightcurves and to produce color-color diagram \citep[e.g.,][]{ZhangGB2011,BakNielsen2017}.
In addition, we note that some X-ray sources close to GRO J1744-28 located in the field of view of \textit{PCA} may contribute to the background.
So we also assumed the background as the real observation when the source was in a very faint state (although it would be overestimated), and found that this only had a slight influence on the following results.
In the hardness-intensity diagram, the hardness gradually increases  with the increasing of the flux when the flux is lower than $\sim$ $\rm 10^{-8}\ ergs\ cm^{-2}\  s^{-1}$.
At a higher flux, the hardness remains almost constant.
This suggests that the spectral shape changes during outbursts, showing a spectral transition around flux $\sim$ $\rm 10^{-8}\ ergs\ cm^{-2}\  s^{-1}$.

\subsection{Spectral analysis}
The broadband X-ray spectrum of GRO J1744-24 has been studied by \citet{Nishiuchi1999,Degenaar2014, Doroshenko2015, DAI2015, Younes2015}.
\citet{Doroshenko2015} described the \textit{BeppoSAX} data of 1997 in the energy range of 1-100\,keV by using typical accreting pulsar models, i.e., a gaussian line plus a power law with a high energy exponential rolloff (gauss + cutoffpl).
\citet{Degenaar2014} used a phenomenological model (a gaussian line + a blackbody + a power law, i.e., "gauss + bbodyrad + powerlaw") to describe the \textit{Chandra} observation in 2014.
\citet{DAI2015} fitted the 1--70\,keV spectrum at the peak of the outburst in 2014 using \textit{XMM-Newton} and \textit{INTEGRAL} observations, and they found that the spectrum can be described with "a gaussian line + a multi temperature blackbody + thermally Comptonised continuum (i.e., gauss+diskbb+nthcomp)" model.
In addition, a CRSF was reported around $\sim$4--5\,keV \citep{Doroshenko2015,DAI2015}.
The energy resolution of \textit{PCA} does not allow, however, to constrain the CRSF parameters or even detect it.
To verify that the influence of the CRSF is negligible, we produced a simulated spectrum by using \textit{PCA}'s response files and the spectral parameters reported in \citet{DAI2015} (around the outburst peak in 2014), in which the CRSF was included. Then, we fitted the simulated spectrum by using the same model but excluding the CRSF. We could not find a clear residual around the line, and the remaining best-fitting parameters were consistent with the input model.
Therefore, in the following analysis, we ignored the contribution from the CRSF.
To take into account the interstellar photo-electric absorption, we used the "wabs" model, in which the hydrogen column density has been fixed to $6.3\times10^{22} {\rm atoms\ {cm}^{-2}}$ \citep{DAI2015}.

\begin{figure}
	\centering
	\includegraphics[width=5.in]{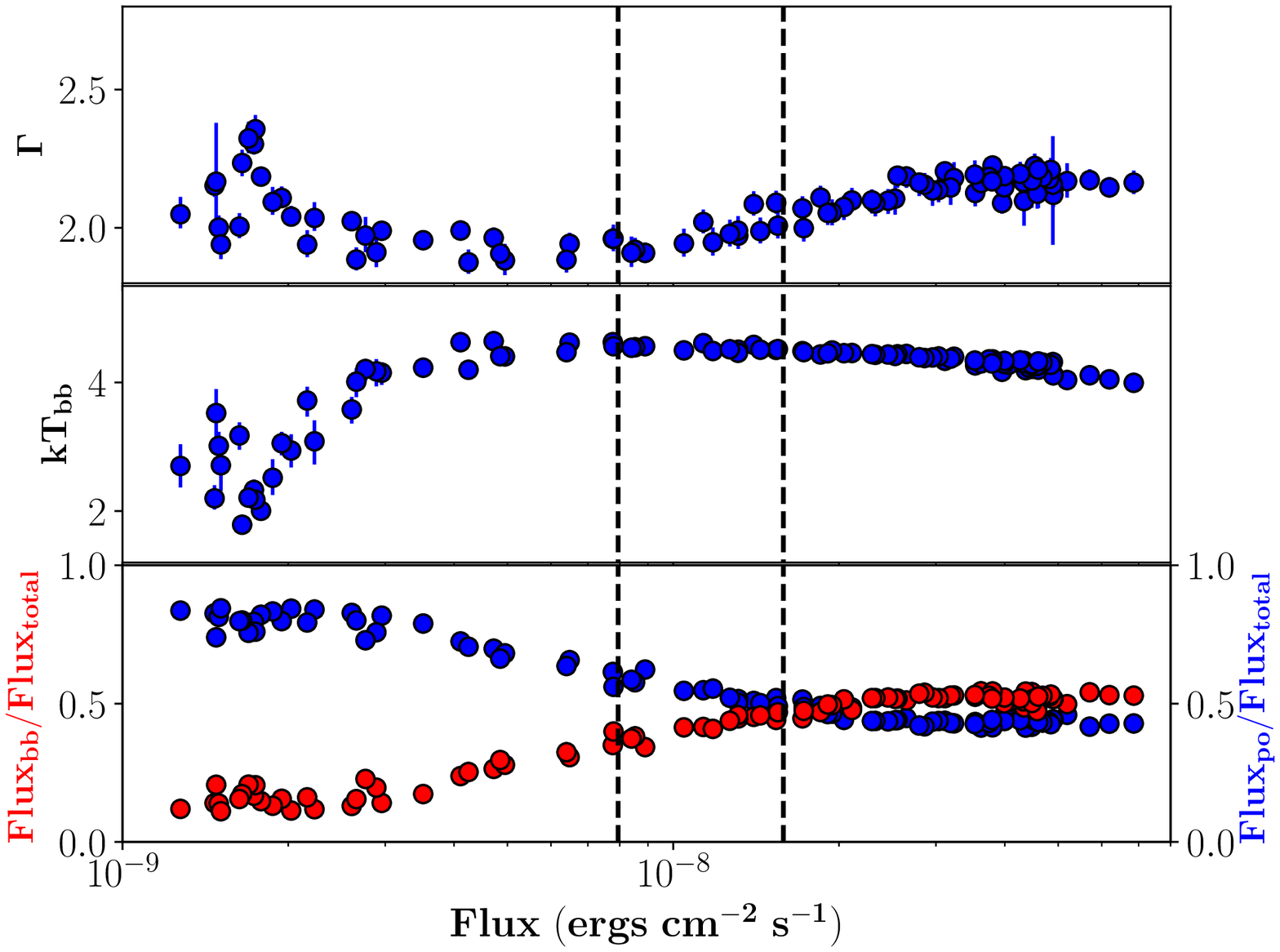}
	\includegraphics[width=5.in]{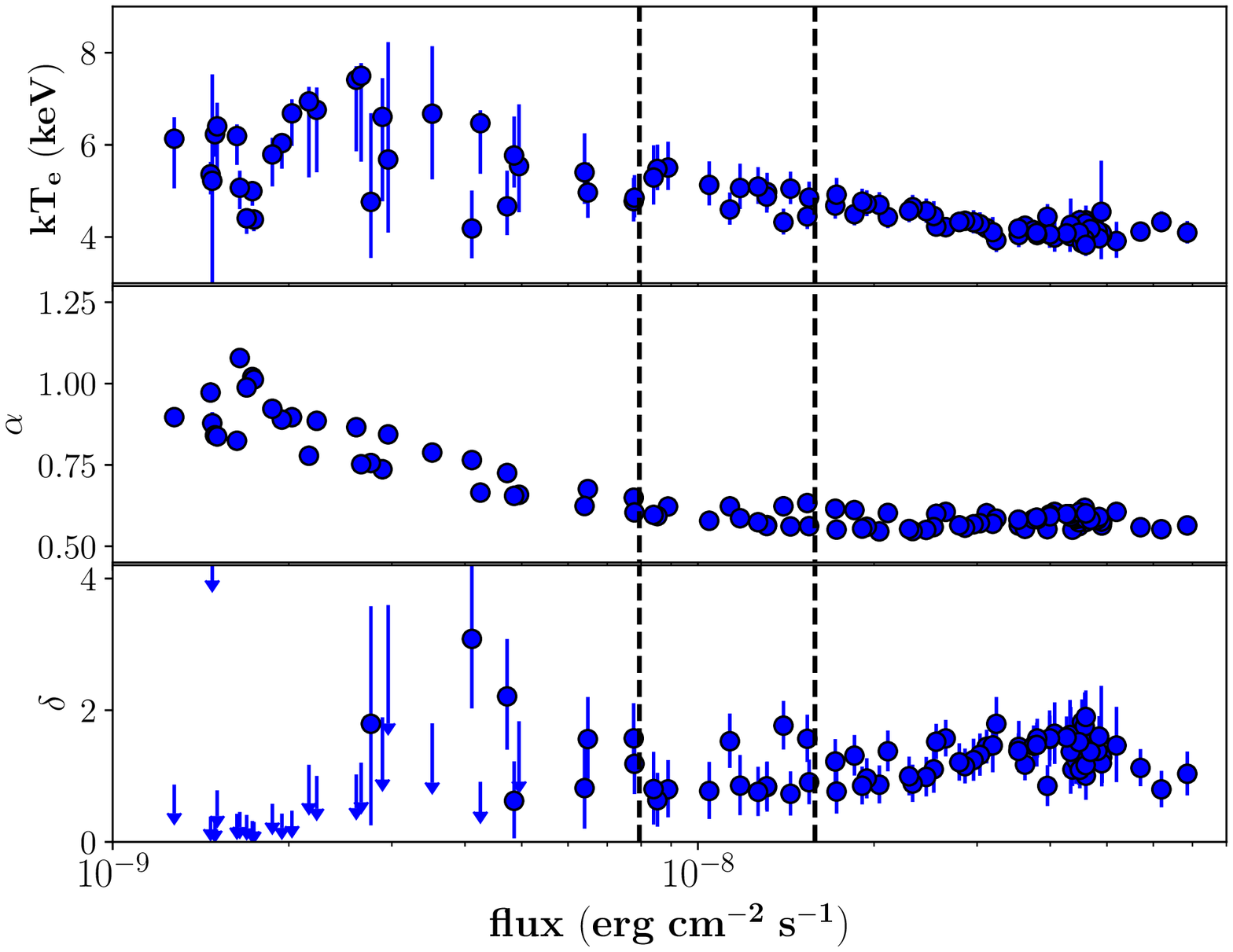}
    \caption{The parameters of spectral fittings by using the wabs*(gauss+powerlaw+bbodyrad) (top panel) and wabs*(gauss + comptb) model (bottom panel).
 	The dashed lines represent the flux equals to 0.8 and 1.6 $\times 10^{-8}$ $\rm ergs\ cm^{-2}\ s^{-1}$, respectively,  corresponding to the transitional flux in the timing analysis.
}
  	\label{pars_flux}
  \end{figure}

\begin{figure}
	\centering
	\includegraphics[width=3.5in, angle=-90]{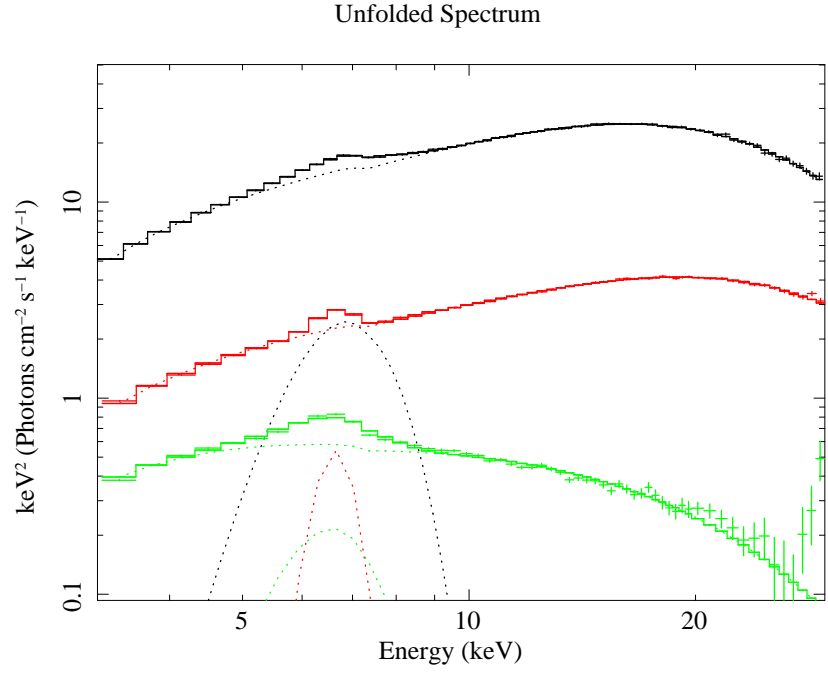}
	\caption{ Examples of unfolded spectra in the bright state (black), around the transitional flux (red), and in the faint state (green), respectively.
	}
	\label{spec}
\end{figure}

We tried to fit all {\it PCA} data with the same model.
First we have tested several phenomenological models typically used to describe spectria of X-ray pulsars, i.e., gauss + bbodyrad + powerlaw, gauss+nthcomp/cutoffpl, and gauss+highecut*powerlaw.
All models are acceptable when the flux is below a few $\rm 10^{-9}\ ergs\ {cm}^{-2}\ s^{-1} $ (3--30\,keV), which may imply that the spectra are dominated by a powerlaw-like/thermal Comptonization component.
On the other hand, only the phenomenological model "wabs*(bbodyrad + powerlaw + gauss)" results in a acceptable ${\chi}^2$ around the outburst peak.
We show the results in Table~\ref{longtable} and Figure~\ref{pars_flux}.
In the faint state ($\rm \lesssim10^{-8}\ ergs\ s^{-1}\ {cm}^{-2}$), the photon index ($\Gamma$) and the blackbody temperature ($kT_{\rm bb}$) seem to respectively decrease and increase with the increasing of the flux, although points are rather scattered.
The flux of the blackbody component increases faster than the power-law component.
The latter dominates the spectra when the source is very faint, and the former becomes gradually important with the increasing of the flux.
In the bright state, the $\Gamma$ and $kT_{\rm bb}$ are relatively constant, and the fluxes of the two components are comparable.
However, as discussed later, we note that the best-fitting parameters of the blackbody component are not consistent with either the accretion disc or the hot spot on the surface of the neutron star, so the interpretation of this component is unclear.
{
In fact the best-fitting parameters of the blackbody components are different from those reported in literature.
This is most likely due to systematic uncertainties arising from the limited response at low energies of the {\it RXTE}/PCA in comparison with the other instruments used in studies found in literature. 
}
We have also used a physical model, i.e., "wabs*(gauss + comptb)", to fit the spectra again, and obtained acceptable results.
The "comptb" model is a hybrid model including thermal and bulk Comptonization \citep{Farinelli2008}, which has been widely used in X-ray millisecond pulsars \citep[e.g.,][]{Farinelli2009, Trumper2010}.
It can describe the Comptonization spectrum of soft photons off electrons which are either purely thermal or additionally subjected to an inward bulk motion.
The model includes 7 free parameters: the temperature of the seed photons ($kT_{\rm s}$), the index of the seed photon spectrum ($\gamma$), the energy index of the Comptonization ($\alpha$), the efficiency of bulk over thermal Comptonization (the bulk parameter, $\delta$), the temperature of the electrons ($kT_{\rm e}$), the illuminating (logA, a factor indicating the importance of the Comptonization compared to the seed photons) and the normalization factor.
For the fits, only some of the parameters could be well constrained, while others had to be frozen.
In this paper, we assume that the seed photons obey a blackbody distribution, and thus that $\gamma$ equals to 3.
We found that the {\it PCA} data was very insensitive to logA and $kT_{\rm s}$, and generally only a upper limit of $kT_{\rm s}$ and a lower limit of logA could be obtained, because {\it PCA} does not have the enough sensitivity below 3\,keV, which is needed to constrain the soft photons not subjected to the Compton scattering.
Therefore we have frozen them at logA=1 and $kT_{\rm s}$=0.3\,keV.
We confirmed that the logA and $kT_{\rm s}$ only had a very little influence on the other parameters.
We show the spectral fitting results in Figure~\ref{pars_flux}.
The $kT_{\rm e}$ parameter is $\sim$ 5--6\,keV when the source is faint (although the scattering is relatively large), and slightly smaller around the outburst peak.
The parameter $\alpha$, instead, decreases from 1 to 0.5 with the increasing flux when the flux is $\lesssim$ $10^{-8}$ ${\rm ergs\ {cm}^{-2}\ s^{-1}}$ , and remains almost unchanged afterwards.
The $\delta$ parameter is $\sim$ 1 when the source is bright, while only an upper limit can be obtained when the source is faint.
It seems that the bulk Comptonization might be only important in the bright state, which is consistent with our conclusion, i.e., the spectra can be well described as a power-law-like component when the flux is low.
We show examples of spectra in bright/transitional/faint states in Figure~\ref{spec}.
They are indeed visibly different.

{In general, we also found a broad emission line at $\sim$ 6.6\,keV, whose width and normalization smoothly increase with the flux.
However, we caution that this effect might be due to changes of the continuum, since the {\it PCA}'s energy resolution does not allow us to well constrain the line feature and its flux in the spectrum.
}
\subsection{Timing analysis}
\begin{figure}
\centering
\includegraphics[width=3.5in, height=6.5in]{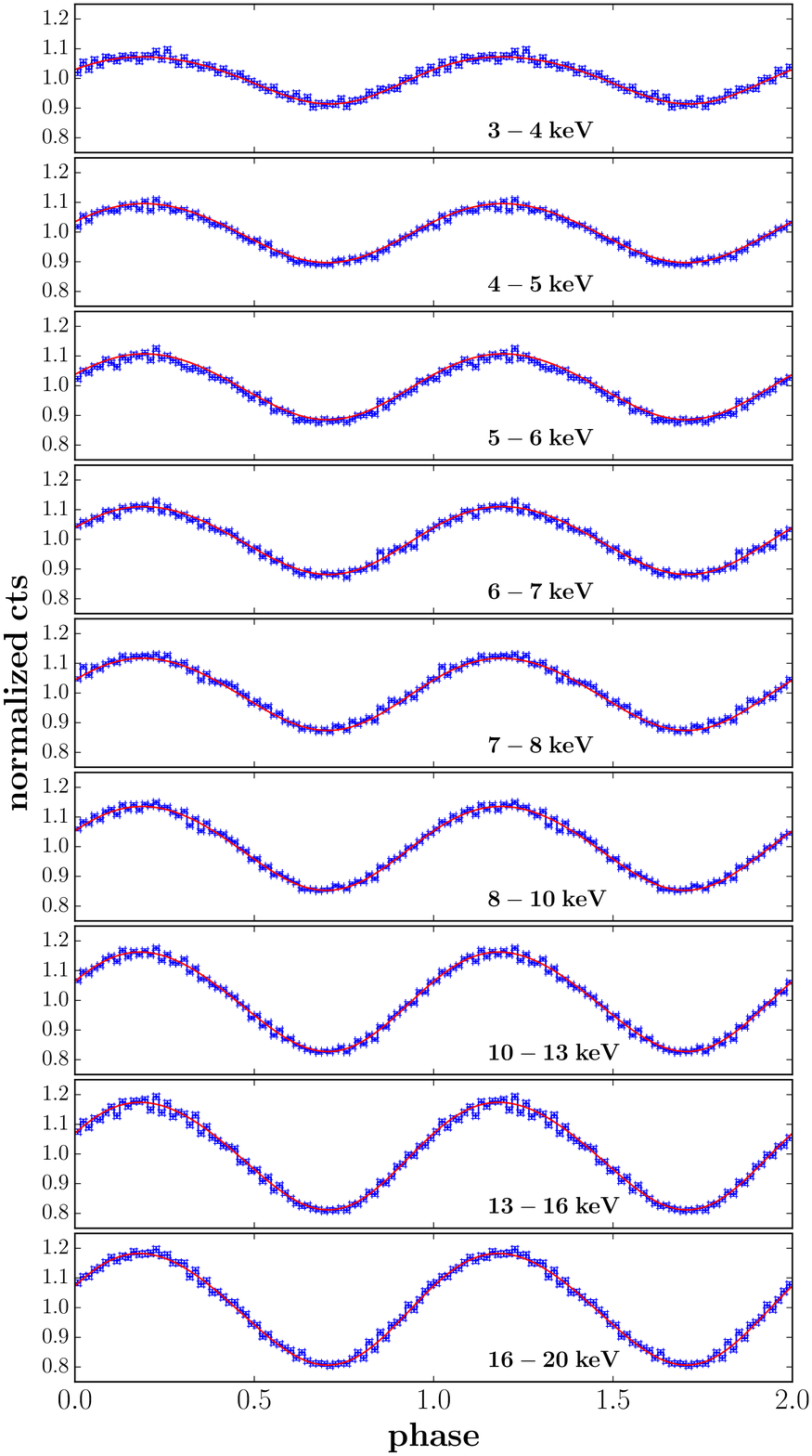}
\caption{An example of folded lightcurves of different energy bands in observational ID 10401-01-04-00, which are fitted by a function of $const + A_1 sin(2\pi \frac{t}{T} + \phi_1) + A_2 sin(4\pi \frac{t}{T} + \phi_2)$.}
\label{fold_example}
\end{figure}

\begin{figure}
\centering
\includegraphics[width=5.5in]{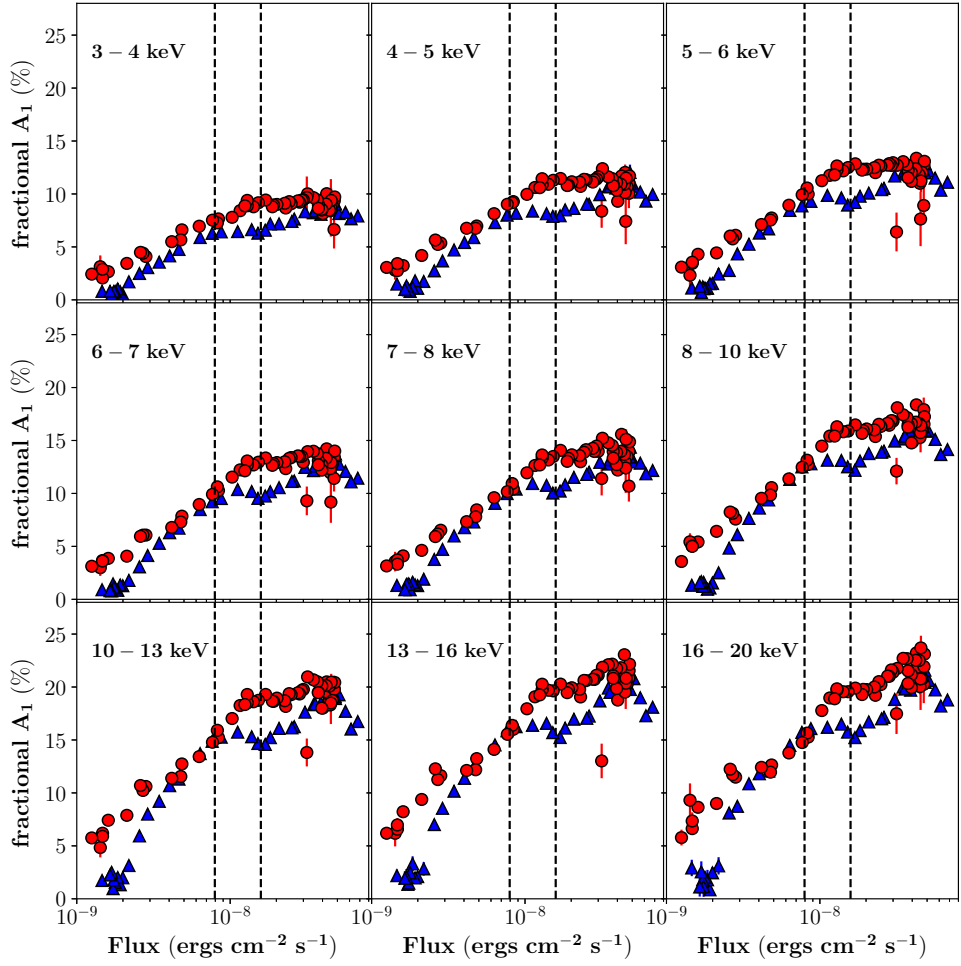}
\caption{
	 The relation between the flux and the fractional pulsed amplitude of $A_1$ in different energy bands.
	 The dashed lines represent the flux equals to 0.8 and 1.6 $\times 10^{-8}$ $\rm ergs\ cm^{-2}\ s^{-1}$, respectively,  corresponding to the transitional flux in the spectral and timing analysis.
}
\label{flux_A1}
\end{figure}

\begin{figure}
\centering
\includegraphics[width=3.2in]{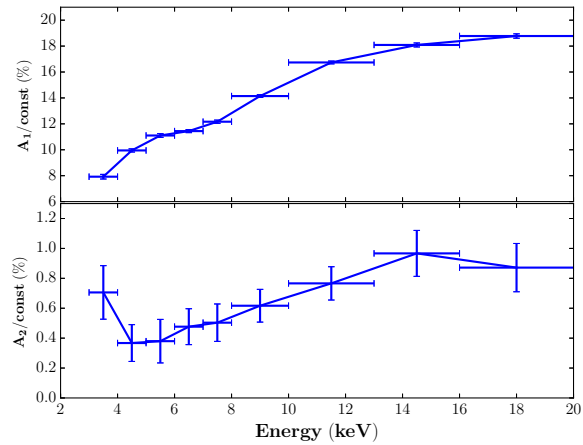}
\caption{An example of the relation between the pulsed amplitude (fundamental and harmonic) and energy around outburst peaks. }
\label{energy_A1}
\end{figure}

In this section, we mainly concentrate on the evolution of the timing properties in different energy bands during outbursts.
As mentioned in Section~I, GRO J1744-28 is a pulsar that has a spin of $\sim$ 2.1\,Hz.
In each observational ID, we estimated its accurate spin frequency by locating the peak of the Rayleigh power ($nR^2$=$\frac{1}{n}{\left|\sum_{i=0}^{n-1}exp\{2\pi\nu t_i\} \right|}^2$), by using the "Event" or "Binned" data in the energy band of 3--30\,keV.
The barycenter correction was carried out by the \textit{Ftools} command \textit{faxbary}.
The ephemeris adopted from \citet{Finger1996} were used to correct the orbital motion of this source.
Then we extracted lightcurves in different energy bands, i.e., 3--4\,keV, 4--5\,keV, 5--6\,keV, 6--7\,keV, 7--8\,keV, 8--10\,keV, 10--13\,keV, 13--16\,keV, 16--20\,keV, and folded them based on the period detected above.
We fitted the pulse profile as a combination of two sinusoids and a constant, i.e., $const + A_1 sin(2\pi \frac{t}{T} + \phi_1) + A_2 sin(4 \pi \frac{t}{T} + \phi_2)$, where the $const$ represents the non-pulsed component, and the two sinusoids are the fundamental and {second} harmonic components of pulsations, and $T$ is the detected period.
Such a function can describe the pulse profile quite well, resulting in an averaged ${\chi}^2$ of 58.93 (dof 59).
We show an example in Figure~\ref{fold_example}.
We note that in most of cases, the {second} harmonic is too weak to be constrained except for a few observations around the outburst peak.
Therefore, unless stated otherwise, we only studied the fundamental component, i.e., $A_1$.
We defined the fractional pulsed amplitude (or pulsed fraction) as the ratio of $A_1$ to $const$.
As shown in Figure~\ref{flux_A1} and~\ref{energy_A1}, the fractional pulsed amplitude of $A_1$ is correlated with both energy and the outburst flux .
In general, $A_1$ is larger in a brighter state and at hard energies.
The $A_1$ parameter at 16--20\,keV, during the outburst peak, reaches $\sim$ 20\%.
In addition, it seems that the positive correlation between the flux and $A_{1}$ is more significant in the faint state.

\begin{figure}
\centering
\includegraphics[width=5.5in]{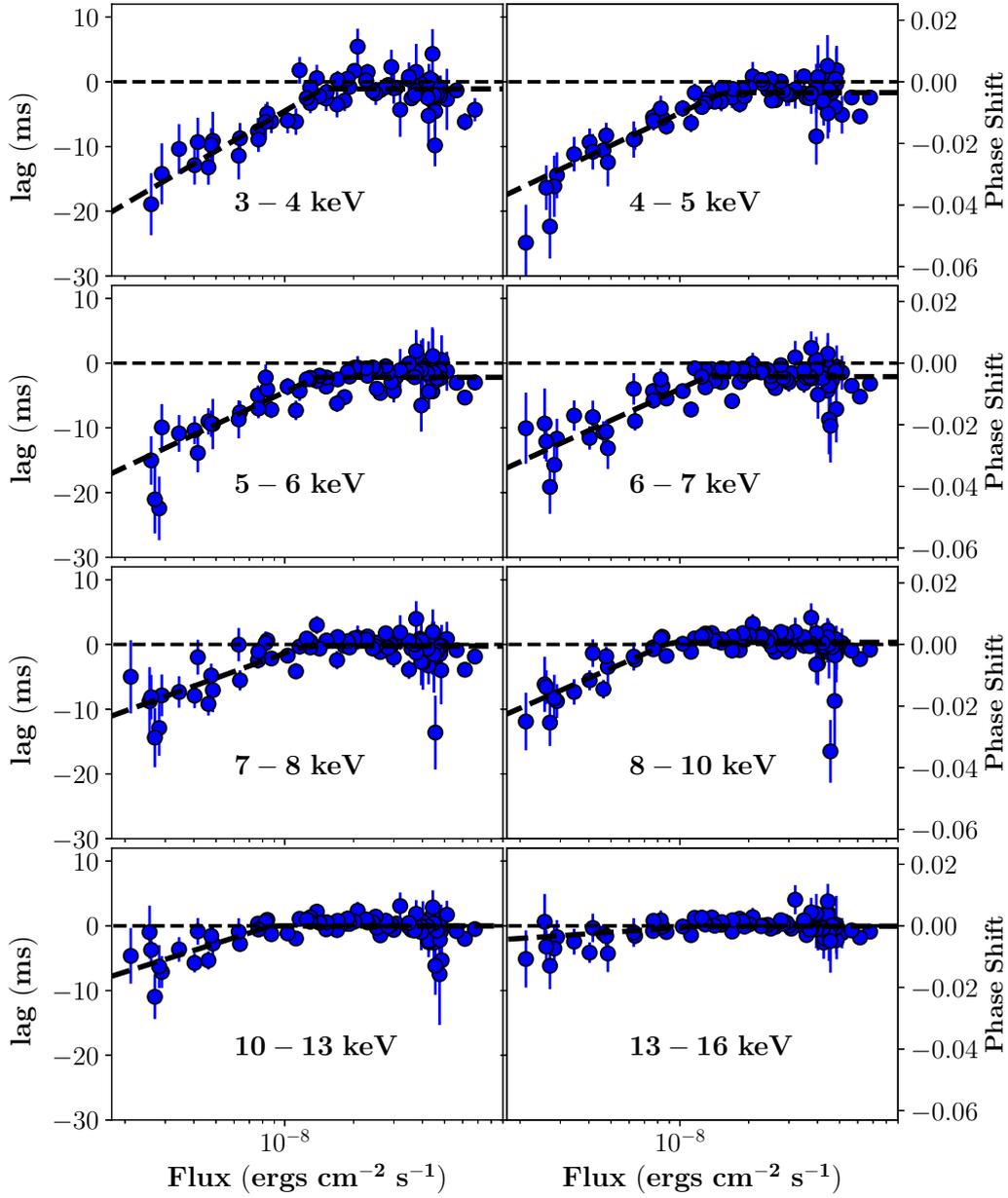}
\caption{
Hard X-ray lags (and the corresponding phase shifts) in different energy bands (with respect to 16--20\,keV) vs.
the flux at 3--30\,keV.
}
\label{flux_lag_13}
\end{figure}

\begin{figure}
\centering
\includegraphics[width=3.2in]{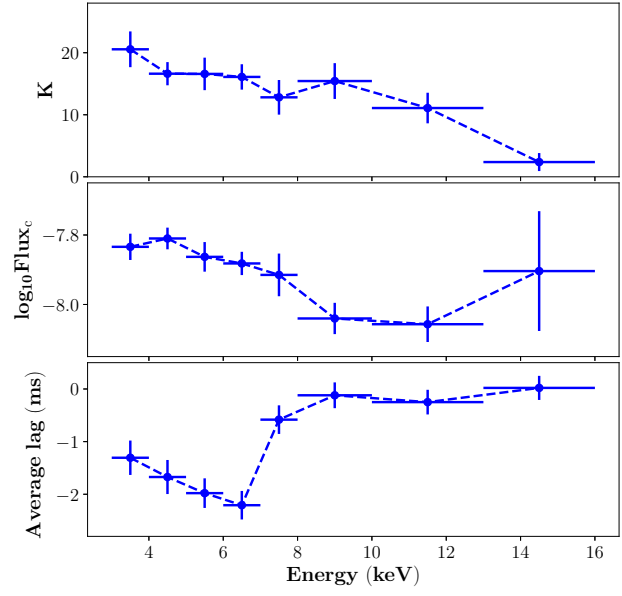}
\caption{
	The upper and middle panels represent slopes (K) and transitional fluxes of the fittings shown in Figure~\ref{flux_lag_13}, respectively.
	The lower panel shows the averaged lags when the flux is larger than the transitional fluxes.
	}
\label{energy_slope}
\end{figure}

For each energy band,  a phase $\phi_1$ can be obtained by fitting the folded lightcurves \citep{DAI2016}.
We studied the phase shift (or the time lag) of soft X-rays compared to the hardest energy band (16--20\,keV) since pulsations detected at hard X-rays are relatively accurate due to a larger $A_1$.
A negative value of the time lag indicates that the soft X-rays precede the hard X-rays.
We show the results in Figure~\ref{flux_lag_13}.
Here we only selected the observations in which the pulsations can be definitely detected, i.e., the ratio of $A_1$ to its error is larger than 5, since the $\phi_1$ is meaningless otherwise.
This criteria is very common in timing analysis \citep[e.g.,][]{BakNielsen2017}.
The lag seems to be larger when the flux is weaker.
We used a function
\begin{eqnarray}
{\rm{lag (flux)}} = \left\{ \begin{array}{l}
{\rm{A}} + {\rm{K}} \times {\log _{10}}{\rm{flux,  flux  <  flu}}{{\rm{x}}_{\rm{c}}}\\
{{\bf constant =} \rm{A}} + {\rm{K}} \times {\log _{10}}{\rm{flu}}{{\rm{x}}_{\rm{c}}}{\rm{, flux }} \ge {\rm{ flu}}{{\rm{x}}_{\rm{c}}}
\end{array} \right.
\end{eqnarray}
to describe the evolution of the hard X-ray lags (see the black dashed lines in Figure~\ref{flux_lag_13}),
where the parameters are the intercept (A), the slope (K) and the transitional flux ($\rm flux_{c}$), respectively.
Such a function can describe the lag evolution quite well, resulting in an averaged ${\chi}^2$ of 121 (82 dof).
We shows the parameters in Figure~\ref{energy_slope}.
Clearly, time lags of hard X-rays are more significant when the source is in a faint state.
Instead, the transitional points ($\rm {flux}_{\rm c}$) are relatively constant, i.e., {in the range 0.8 to 1.6 $\rm  \times{10}^{-8}\ ergs\ s^{-1}\ {cm}^{-2}$}, for different energy bands (see middle panel of  Figure~\ref{energy_slope}).
We note that this flux range exactly corresponds to the spectral transition (see the dashed lines in Figure~\ref{lightcurve_and_HID},\ref{pars_flux}).
This indicates that both the timing and spectral transitions arise from the same physical process.
When the flux is larger than the transitional flux, the lags is almost independent of the flux.
In the lower panel of Figure~\ref{energy_slope} we show the averaged lags for different energy bands, when the flux is larger than $\rm flux_c$.
In this case, the phase shift between hard and soft X-rays is significantly weaker.

\section{Discussion and conclusions}
We systematically analyzed {\it RXTE/PCA} data to investigate the spectral and timing evolutions along two outbursts of GRO J1744-28.
Their lightcurves are quite similar, showing a duration of a hundred days and a peak flux of $\rm \sim 5\times {10}^{-8}\,ergs\ {cm}^{-2}\ {s}^{-1}$ (3--30\,keV).
From the intensity-hardness diagram, we have found a clear spectral  transition around $\rm \sim10^{-8}\,ergs\ {cm}^{-2}\ s^{-1}$.
Therefore, in this paper, we define the bright (faint) state as the flux larger (smaller) than this threshold.
In the faint state, the hardness is positively correlated to the flux, while remains rather constant in the bright state.
The transitional luminosity is $\sim$ 7.6 or 1.9 $\rm \times 10^{37} ergs\ s^{-1}$ assuming a distance of 8\,kpc or 4\,kpc (corresponding to the study of the absorption and the accretion torque \citep{Kouveliotou1996, Sanna2017}, respectively) and an isotropic radiation.
Similar transitions have been reported in Be/X-ray pulsars by \citet{Reig2013}.
In these systems, however, the hardness is anti-correlated with the flux in the bright state instead of being independent.

The spectra of GRO J1744-28 can not be well fitted by a simple model, e.g., cutoffpl or highecut*powerlaw, especially when the source is in the bright state.
A phenomenological model of "wabs*(gauss+bbodyrad + powerlaw)" can describe the spectra successfully, which is fully consistent with previous reports \citep[e.g.,][]{Degenaar2014}.
{
The powerlaw component dominates the spectrum when the source is in the faint state, while the blackbody component is gradually important with the increasing luminosity.
The temperature of the blackbody component is $\sim$ 2--4\,keV, higher than values found in previous observations, i.e., $\sim$ 0.5-1\,keV \citep{Degenaar2014, Younes2015}.
We suggest that this might be due to systematic uncertainties arising from the different energy ranges observed by different instruments. 
However, if we assume that the blackbody component originates from a hot spot on the surface of the neutron star with a distance of 8\,kpc, the inferred radius increases smoothly from 0.6 to 6\,km with the increasing luminosity.
This value is consistent with the size of the polar cap ($\sim$1\,km) only if the source is relatively faint.
On the other hand, the accretion disc, which is truncated far away from the neutron star, can not contribute to half of the total energy budget in the bright state.
A high energy cutoff ($E_{\rm cut}$) around $\sim$ 18\,keV was reported by, e.g., \citet{Doroshenko2015}, using the broad band capabilities of {\it BeppoSAX}.
However, the $E_{\rm cut}$ could not be well constrained by using our data.
We compared the simple powerlaw model with a {\it cutoffpl} model by using an F-test, and found that on average the latter could improve fits only at $\sim 1 \sigma$ confidence level.
}
We also tried to fit the spectra by using a thermal and bulk Comptonization model ({\it comptb}), and obtained acceptable results.
In this case, the temperature of the hot electrons is $\sim$ 6\,keV in the faint state, and slightly decreases in the bright state.
The energy index $\alpha$ decreases from 1\ to 0.5 when the flux is $\rm \lesssim 10^{-8}\ erg\ {cm}^{-2}\ s^{-1}$, and remains almost unchanged in the brighter state.
The parameter $\alpha$ is a free parameter when calculating the Green function of the
Comptonization spectrum \citep{Farinelli2008}, and can approximately be used to infer the optical depth ($\tau$) by using the following relation:
\[\alpha  =  - \frac{3}{2} + \sqrt {\frac{9}{4} + \frac{{{\pi ^2}{{\rm{m}}_{\rm{e}}}{{\rm{c}}^2}}}{{\eta {\rm{k}}{{\rm{T}}_{\rm{e}}}{{{\rm{(}}\tau {\rm{ + }}\frac{2}{3}{\rm{)}}}^2}}}} ,\]
where $\eta$ represents the spherical ($\eta$=3) , plane ($\eta$=12) or cylindrical ($\eta$=16/3) geometry \citep{Titarchuk1997, Sunyaev1980}.
Therefore, the decreasing $\alpha$ implies an increase of $\tau$ from 7.7 to 13.2 (spherical geometry ), from 3.5 to 6.3 (plane geometry) or from 5.6 to 9.7 (cylindrical geometry).
We note that the evolution of  $\alpha$ corresponds to the transition in the intensity-hardness diagram, i.e., both $\alpha$ and the hardness evolve significantly in the faint state and remain unchanged in the bright state (see Figure \ref{lightcurve_and_HID}, \ref{pars_flux}).
The bulk parameter $\delta$ is $\sim$ 1 in the bright state, while sometimes only an upper limit can be constrained in the faint state.
It seems that the $\delta$ is smaller in the faint state, although the error bars are quite large.
This suggests that the bulk Comptonization might be more significant in the bright state.

In the timing analysis, we find that the pulsed fraction is positively correlated to both the energy band (at 3--30\,keV) and luminosity.
This result is well in agreement with the reports by \citet{DAI2015, Doroshenko2015,Younes2015, DAI2016, Cui1997}.
In addition, thanks to the stable pulse profiles in different energy bands, we studied the energy-dependent lags with respect to the 16--20\,keV band.
We find that the lag (the soft X-rays precede the hard X-rays) is more significant when the flux is $\lesssim$ $\rm  {10}^{-8}\ ergs\ s^{-1}\ {cm}^{-2}$.
The hard X-ray lags depend on the energy band, which is consistent with the results reported by \citet{DAI2016}.
The drop in the lag spectrum around 6.4\,keV is not significant (see the upper panel in Figure~\ref{energy_slope}) based on {\it PCA} observations, which is likely due to its poor energy resolution.
The lag between 3--4\,keV and 16--20\,keV is up to $\sim$ 20\,ms in the faint state.
However, the lag is much smaller ($\lesssim$ 2\,ms) in the bright state.

It has been suggested that the hard X-ray lag is due to the Compton reverberation-like process \citep{Lightman1978, Payne1980, Guilbert1982, DAI2016}.
In a nutshell,  in a hot, thermal plasma hard X-rays can be produced by repeated Compton scattering, because the energy of the incident photons can be amplified by a factor of $\sim$ 1+$\frac{4kT}{m_{\rm e}c^2}$ for non-relativistic electrons per scattering, where the kT is the temperature of the hot plasma.
In this case, the harder photons need more scattering times,  and then lead to more delays.
In theory, the hard X-ray lags depend on both the optical depth ($\tau$) and the Compton region's size ($l$).
Therefore, the suppressed hard X-ray lags in the bright state might be due to the increasing $\tau$ or the decreasing $l$.
By assuming a Comptb model, we found an increasing of $\tau$ by a factor of 2, which may qualitatively explain the reduced hard X-ray lags.
However, we note that the energy budget is a big issue for the Compton scattering model.
Because the soft photons have to gain energy from the Comptonization cloud though up-scatterings, high temperature plasma is required.
While the temperature of electrons obtained from the spectral analysis is quite low (a few keV), which cannot boost the seed photons to a few tens of keV.
On the other hand, we speculate that the hard X-ray lags might originate from a geometry effect.
In this scenario, the soft and hard X-rays are actually independently emitted in different angular distribution.
The maximum hard X-ray lag (20\,ms, between 3--4 and 16--20\,keV) corresponds to a phase shift of $\sim$ 4\%, which means a projection distance of $\sim$ 2.5\,km on the neutron star by assuming a 10\,km neutron star radius.
A simple configuration is that the photons escaping from the accretion column may irradiate and heat the surface of the neutron star \citep[see, e.g., Fig. 1 in][]{Poutanen2013}.
The reflection region may be a few kilometers away from the polar cap, and result in this phase shift.
The reflection region is predicted to be visibly not axis-symmetric, otherwise the phase shift will be cancelled out.
However, our conjecture to be proven true, requires more detailed physical models, which are still missing.

What is interesting is that the flux at which the transition of the timing property is observed, i.e., the hard X-ray lags,  exactly equals the critical flux of the spectral transition.
This means that the timing and spectral transitions actually could arise from same physical processes.
The morphology of the accreting structure onto the neutron star surface depends on luminosity.
If the luminosity is larger than a critical value $L_{\rm c}$,
the infalling matter will be stopped by a radiation-dominated shock and then sinks onto the neutron star surface.
On the contrary, if the luminosity is smaller than the $L_{c}$, the accreted matter can free fall almost down to the surface of the neutron star, and form an accretion mound \citep{Basko1975, Basko1976} .
The transition flux between the two accreting regimes is $L=2.72\times {10}^{37} \left( {\frac{{{\sigma _T}}}{{\sqrt {{\sigma _\parallel }\sigma  \bot } }}} \right)\left( {\frac{{{r_0}}}{R}} \right)\left( {\frac{M}{{M_{\odot}}}} \right) \sim {\rm {10}^{36} - {10}^{37}\,ergs/s}$ \citep{Basko1976, Becker1998, Becker2012, Mushtukov2015}.
Assuming a distance of 4\,kpc, this value is marginally consistent with the transitional luminosity of the spectral and timing properties that we found.
It is, however, too small if the source's distance is 8\,kpc.
Another possibility could be that this transitional luminosity might derive from two different states of the accretion disc \citep{Mushtukov2015b}.
The accretion disc is dominated by the gas and radiation pressure at the low and high accretion rates, respectively.
\citet{Suleimanov2007} proposed that the geometry of the accretion disc at the magnetospheric radius may have an influence on the accretion structure on the neutron star.
Therefore, the transition between the two disc states may lead to the spectral and timing transitions in GRO J1744-28.
As shown in \citet[][Figure 4]{Mushtukov2015b}, for a magnetic field of $\sim 10^{11}$\,G, the transitional flux for disc states is $\rm \sim10^{38}\,ergs\ s^{-1} $, which is consistent with our observations assuming a distance of 8\,kpc.
In addition, based on this model a maximum luminosity of a few $\rm 10^{38}\,ergs\ s^{-1} $  is predicted for low magnetic fields, which is also consistent with the largest luminosity around the outburst peaks of GRO J1744-28.
The transition of the accretion structure on the surface of the neutron can naturally explain the changes of the spectral and timing properties, although detailed theoretical models are still missing.

\section*{Acknowledgements}
JL thanks the support from the Chinese NSFC 11733009.
ZS thanks the support from XTP project XDA 04060604, the Strategic Priority Research Programme 'The Emergence of Cosmological Structures' of the Chinese Academy of Sciences, Grant No.XDB09000000,the National Key Research and Development Program of China (2016YFA0400800) and the Chinese NSFC 11473027 and 11733009.
%
VFS thanks the support from the German Research Foundation (DFG) grant WE 1312/51-1 and the Russian Government Program of Competitive Growth of Kazan Federal University.

\bibliographystyle{mnras}
\bibliography{mybibtex}

\clearpage
\appendix
 {
	\begin{longtable}{c c | c c | c c c | c}
		\hline
		& & \multicolumn{2}{|c|}{bbodyrad + powerlaw} &
		\multicolumn{3}{|c|}{Comptb}                   &  \\
		ObsId               & Time  & Powerlaw Index                    &  $kT_{\rm bb}$                    &
		$\alpha$                & $kT_{\rm comptb}$       &      $\delta$                    & Flux (3--30\,keV) \\
		&  (MJD)    &                 &        (keV)           &            &  (keV)      &      & ($\rm 10^{-9}\,ergs\ s^{-1}\, {cm}^{-2}$)  \\
		\hline
		10401-01-04-00  &  50128.29 &  $2.16_{-0.04}^{+0.04}$  &  $3.99_{-0.03}^{+0.03}$ & $0.56_{-0.01}^{+0.01}$  &  $4.10_{-0.23}^{+0.25}$ &   $1.04_{-0.33}^{+0.34}$ &  68.59 \\
		10401-01-05-00  &  50129.47 &  $2.17_{-0.06}^{+0.06}$  &  $4.04_{-0.05}^{+0.05}$ & $0.61_{-0.01}^{+0.01}$  &  $3.91_{-0.36}^{+0.42}$ &   $1.46_{-0.56}^{+0.59}$ &  51.92 \\
		10401-01-06-00  &  50134.54 &  $2.15_{-0.04}^{+0.04}$  &  $4.05_{-0.03}^{+0.03}$ & $0.55_{-0.01}^{+0.01}$  &  $4.33_{-0.21}^{+0.23}$ &   $0.80_{-0.27}^{+0.28}$ &  61.95 \\
		10401-01-09-00  &  50138.82 &  $2.17_{-0.04}^{+0.04}$  &  $4.11_{-0.03}^{+0.03}$ & $0.56_{-0.01}^{+0.01}$  &  $4.12_{-0.20}^{+0.21}$ &   $1.12_{-0.28}^{+0.29}$ &  57.06 \\
		10401-01-08-00  &  50142.16 &  $2.14_{-0.05}^{+0.05}$  &  $4.12_{-0.04}^{+0.04}$ & $0.57_{-0.01}^{+0.01}$  &  $4.05_{-0.27}^{+0.30}$ &   $1.28_{-0.39}^{+0.40}$ &  49.05 \\
		10401-01-10-00  &  50143.89 &  $2.12_{-0.04}^{+0.05}$  &  $4.11_{-0.04}^{+0.03}$ & $0.56_{-0.01}^{+0.01}$  &  $4.09_{-0.25}^{+0.28}$ &   $1.19_{-0.35}^{+0.36}$ &  49.00 \\
		10401-01-11-00  &  50148.02 &  $2.17_{-0.04}^{+0.04}$  &  $4.21_{-0.03}^{+0.03}$ & $0.56_{-0.01}^{+0.01}$  &  $4.01_{-0.20}^{+0.21}$ &   $1.39_{-0.29}^{+0.30}$ &  44.80 \\
		10401-01-12-00  &  50151.89 &  $2.16_{-0.04}^{+0.04}$  &  $4.19_{-0.03}^{+0.03}$ & $0.55_{-0.01}^{+0.01}$  &  $4.20_{-0.18}^{+0.20}$ &   $1.10_{-0.25}^{+0.26}$ &  43.66 \\
		10401-01-13-00  &  50155.90 &  $2.09_{-0.04}^{+0.04}$  &  $4.17_{-0.04}^{+0.03}$ & $0.55_{-0.01}^{+0.01}$  &  $4.44_{-0.25}^{+0.27}$ &   $0.85_{-0.30}^{+0.31}$ &  39.56 \\
		10401-01-16-00  &  50156.76 &  $2.13_{-0.05}^{+0.05}$  &  $4.27_{-0.04}^{+0.04}$ & $0.56_{-0.01}^{+0.01}$  &  $4.04_{-0.26}^{+0.29}$ &   $1.44_{-0.38}^{+0.39}$ &  35.35 \\
		10401-01-14-00  &  50159.03 &  $2.16_{-0.03}^{+0.04}$  &  $4.30_{-0.03}^{+0.02}$ & $0.55_{-0.01}^{+0.01}$  &  $4.25_{-0.17}^{+0.18}$ &   $1.17_{-0.24}^{+0.24}$ &  36.23 \\
		10401-01-17-00  &  50164.53 &  $2.20_{-0.03}^{+0.03}$  &  $4.34_{-0.03}^{+0.03}$ & $0.60_{-0.01}^{+0.01}$  &  $4.19_{-0.17}^{+0.19}$ &   $1.44_{-0.25}^{+0.26}$ &  31.14 \\
		10401-01-18-00  &  50171.81 &  $2.19_{-0.04}^{+0.04}$  &  $4.44_{-0.03}^{+0.02}$ & $0.61_{-0.01}^{+0.01}$  &  $4.22_{-0.19}^{+0.20}$ &   $1.57_{-0.27}^{+0.28}$ &  26.52 \\
		10401-01-19-00  &  50172.73 &  $2.19_{-0.04}^{+0.02}$  &  $4.44_{-0.04}^{+0.01}$ & $0.60_{-0.01}^{+0.01}$  &  $4.23_{-0.19}^{+0.20}$ &   $1.52_{-0.26}^{+0.27}$ &  25.55 \\
		10401-01-20-00  &  50178.94 &  $2.10_{-0.04}^{+0.05}$  &  $4.46_{-0.04}^{+0.04}$ & $0.60_{-0.01}^{+0.01}$  &  $4.43_{-0.24}^{+0.26}$ &   $1.38_{-0.30}^{+0.31}$ &  21.12 \\
		10401-01-21-00  &  50181.53 &  $2.11_{-0.04}^{+0.04}$  &  $4.43_{-0.04}^{+0.04}$ & $0.61_{-0.01}^{+0.01}$  &  $4.50_{-0.24}^{+0.26}$ &   $1.31_{-0.30}^{+0.31}$ &  18.51 \\
		10401-01-22-00  &  50183.40 &  $2.07_{-0.04}^{+0.04}$  &  $4.49_{-0.05}^{+0.05}$ & $0.62_{-0.01}^{+0.01}$  &  $4.68_{-0.28}^{+0.31}$ &   $1.22_{-0.33}^{+0.34}$ &  17.17 \\
		10401-01-22-01  &  50185.62 &  $2.09_{-0.04}^{+0.04}$  &  $4.51_{-0.05}^{+0.04}$ & $0.63_{-0.01}^{+0.01}$  &  $4.45_{-0.27}^{+0.30}$ &   $1.57_{-0.35}^{+0.36}$ &  15.38 \\
		10401-01-23-00  &  50187.67 &  $2.09_{-0.04}^{+0.05}$  &  $4.58_{-0.04}^{+0.04}$ & $0.62_{-0.01}^{+0.01}$  &  $4.32_{-0.27}^{+0.29}$ &   $1.77_{-0.36}^{+0.37}$ &  14.00 \\
		10401-01-24-00  &  50189.88 &  $2.02_{-0.04}^{+0.05}$  &  $4.61_{-0.05}^{+0.05}$ & $0.62_{-0.01}^{+0.01}$  &  $4.60_{-0.33}^{+0.37}$ &   $1.53_{-0.40}^{+0.42}$ &  11.34 \\
		10401-01-25-00  &  50192.63 &  $1.91_{-0.03}^{+0.04}$  &  $4.56_{-0.07}^{+0.07}$ & $0.62_{-0.01}^{+0.01}$  &  $5.51_{-0.49}^{+0.56}$ &   $0.80_{-0.43}^{+0.44}$ &  8.89 \\
		10401-01-26-00  &  50194.69 &  $1.96_{-0.04}^{+0.04}$  &  $4.63_{-0.07}^{+0.06}$ & $0.65_{-0.01}^{+0.01}$  &  $4.78_{-0.44}^{+0.51}$ &   $1.57_{-0.51}^{+0.53}$ &  7.77 \\
		10401-01-27-00  &  50196.62 &  $1.94_{-0.04}^{+0.04}$  &  $4.62_{-0.08}^{+0.07}$ & $0.68_{-0.01}^{+0.01}$  &  $4.97_{-0.55}^{+0.65}$ &   $1.56_{-0.60}^{+0.64}$ &  6.49 \\
		10401-01-28-00  &  50199.95 &  $1.96_{-0.04}^{+0.04}$  &  $4.64_{-0.09}^{+0.08}$ & $0.73_{-0.01}^{+0.01}$  &  $4.67_{-0.62}^{+0.77}$ &   $2.21_{-0.81}^{+0.87}$ &  4.72 \\
		10401-01-29-00  &  50201.69 &  $1.99_{-0.03}^{+0.04}$  &  $4.63_{-0.10}^{+0.09}$ & $0.76_{-0.01}^{+0.01}$  &  $4.19_{-0.64}^{+0.82}$ &   $3.08_{-1.05}^{+1.16}$ &  4.11 \\
		10401-01-30-00  &  50203.82 &  $1.96_{-0.03}^{+0.03}$  &  $4.23_{-0.15}^{+0.13}$ & $0.79_{-0.01}^{+0.01}$  &  $6.68_{-1.42}^{+1.46}$ &   $<1.80$                &  3.52 \\
		10401-01-31-00  &  50206.76 &  $1.99_{-0.03}^{+0.04}$  &  $4.15_{-0.19}^{+0.16}$ & $0.84_{-0.01}^{+0.01}$  &  $5.68_{-1.58}^{+2.54}$ &   $1.65_{-1.61}^{+1.94}$ &  2.96 \\
		10401-01-32-00  &  50209.49 &  $2.02_{-0.03}^{+0.04}$  &  $3.58_{-0.22}^{+0.19}$ & $0.87_{-0.01}^{+0.01}$  &  $7.41_{-1.55}^{+0.29}$ &   $<1.02$                &  2.61 \\
		10401-01-34-00  &  50213.94 &  $2.04_{-0.05}^{+0.06}$  &  $3.08_{-0.36}^{+0.33}$ & $0.89_{-0.01}^{+0.01}$  &  $6.76_{-1.35}^{+0.48}$ &   $<1.00$                &  2.23 \\
		10401-01-37-00  &  50215.66 &  $2.04_{-0.03}^{+0.04}$  &  $2.94_{-0.26}^{+0.25}$ & $0.90_{-0.01}^{+0.01}$  &  $6.68_{-0.71}^{+0.30}$ &   $<0.47$                &  2.03 \\
		10401-01-35-00  &  50216.42 &  $2.11_{-0.04}^{+0.04}$  &  $3.05_{-0.19}^{+0.18}$ & $0.89_{-0.01}^{+0.01}$  &  $6.04_{-0.55}^{+0.21}$ &   $<0.43$                &  1.95 \\
		10401-01-38-00  &  50217.32 &  $2.09_{-0.05}^{+0.05}$  &  $2.52_{-0.27}^{+0.29}$ & $0.92_{-0.02}^{+0.02}$  &  $5.79_{-0.69}^{+0.36}$ &   $<0.58$                &  1.87 \\
		10401-01-39-00  &  50218.84 &  $2.19_{-0.03}^{+0.03}$  &  $2.00_{-0.11}^{+0.13}$ & $1.02_{-0.01}^{+0.01}$  &  $5.43_{-0.22}^{+0.24}$ &   $<0.22$                &  1.78 \\
		10401-01-36-00  &  50219.45 &  $2.30_{-0.04}^{+0.04}$  &  $2.33_{-0.15}^{+0.16}$ & $1.02_{-0.01}^{+0.01}$  &  $5.00_{-0.32}^{+0.17}$ &   $<0.32$                &  1.73 \\
		10401-01-40-00  &  50220.26 &  $2.36_{-0.05}^{+0.05}$  &  $2.18_{-0.13}^{+0.15}$ & $1.01_{-0.02}^{+0.02}$  &  $4.38_{-0.26}^{+0.08}$ &   $<0.30$                &  1.74 \\
		10401-01-41-00  &  50221.27 &  $2.32_{-0.05}^{+0.06}$  &  $2.21_{-0.15}^{+0.17}$ & $0.99_{-0.01}^{+0.02}$  &  $4.41_{-0.34}^{+0.21}$ &   $<0.41$                &  1.69 \\
		10401-01-42-00  &  50224.60 &  $2.15_{-0.04}^{+0.04}$  &  $2.20_{-0.17}^{+0.20}$ & $0.97_{-0.02}^{+0.02}$  &  $5.36_{-0.43}^{+0.27}$ &   $<0.38$                &  1.47 \\
		10401-01-43-00  &  50225.95 &  $2.23_{-0.05}^{+0.05}$  &  $1.79_{-0.10}^{+0.13}$ & $1.08_{-0.02}^{+0.02}$  &  $5.07_{-0.46}^{+0.37}$ &   $<0.45$                &  1.65 \\
		20078-01-03-00  &  50466.01 &  $2.14_{-0.06}^{+0.04}$  &  $4.20_{-0.05}^{+0.03}$ & $0.57_{-0.01}^{+0.01}$  &  $4.36_{-0.29}^{+0.32}$ &   $1.01_{-0.37}^{+0.38}$ &  46.02 \\
		20078-01-03-01  &  50466.11 &  $2.14_{-0.04}^{+0.04}$  &  $4.24_{-0.03}^{+0.03}$ & $0.57_{-0.01}^{+0.01}$  &  $4.26_{-0.23}^{+0.25}$ &   $1.17_{-0.30}^{+0.31}$ &  44.79 \\
		20078-01-03-02  &  50466.21 &  $2.15_{-0.05}^{+0.04}$  &  $4.23_{-0.04}^{+0.03}$ & $0.58_{-0.01}^{+0.01}$  &  $4.21_{-0.26}^{+0.29}$ &   $1.24_{-0.36}^{+0.37}$ &  44.30 \\
		20077-01-01-00  &  50467.88 &  $2.17_{-0.05}^{+0.03}$  &  $4.25_{-0.04}^{+0.03}$ & $0.59_{-0.01}^{+0.01}$  &  $4.36_{-0.21}^{+0.23}$ &   $1.10_{-0.28}^{+0.29}$ &  44.95 \\
		20077-01-02-00  &  50468.94 &  $2.13_{-0.06}^{+0.06}$  &  $4.27_{-0.05}^{+0.04}$ & $0.59_{-0.01}^{+0.01}$  &  $4.07_{-0.32}^{+0.36}$ &   $1.51_{-0.46}^{+0.48}$ &  44.77 \\
		20077-01-03-00  &  50469.94 &  $2.14_{-0.05}^{+0.05}$  &  $4.28_{-0.04}^{+0.04}$ & $0.59_{-0.01}^{+0.01}$  &  $3.98_{-0.27}^{+0.29}$ &   $1.62_{-0.40}^{+0.41}$ &  45.39 \\
		20077-01-04-00  &  50471.00 &  $2.16_{-0.06}^{+0.03}$  &  $4.26_{-0.05}^{+0.03}$ & $0.58_{-0.01}^{+0.01}$  &  $4.30_{-0.25}^{+0.27}$ &   $1.17_{-0.32}^{+0.33}$ &  45.99 \\
		20077-01-05-00  &  50472.22 &  $2.10_{-0.09}^{+0.07}$  &  $4.34_{-0.09}^{+0.05}$ & $0.59_{-0.01}^{+0.01}$  &  $4.26_{-0.48}^{+0.57}$ &   $1.37_{-0.64}^{+0.68}$ &  43.36 \\
		20077-01-06-00  &  50473.14 &  $2.16_{-0.22}^{+0.18}$  &  $4.32_{-0.22}^{+0.10}$ & $0.58_{-0.03}^{+0.02}$  &  $4.55_{-1.03}^{+1.11}$ &   $<2.37$                &  48.92 \\
		20077-01-07-00  &  50473.87 &  $2.18_{-0.07}^{+0.03}$  &  $4.28_{-0.06}^{+0.03}$ & $0.61_{-0.01}^{+0.01}$  &  $3.99_{-0.31}^{+0.34}$ &   $1.64_{-0.46}^{+0.48}$ &  40.70 \\
		20077-01-08-00  &  50475.08 &  $2.17_{-0.08}^{+0.05}$  &  $4.30_{-0.06}^{+0.04}$ & $0.60_{-0.01}^{+0.01}$  &  $4.02_{-0.34}^{+0.39}$ &   $1.62_{-0.51}^{+0.53}$ &  43.24 \\
		20078-01-04-02  &  50476.09 &  $2.18_{-0.04}^{+0.04}$  &  $4.28_{-0.03}^{+0.03}$ & $0.58_{-0.01}^{+0.01}$  &  $4.13_{-0.21}^{+0.22}$ &   $1.38_{-0.30}^{+0.31}$ &  48.19 \\
		20078-01-04-01  &  50477.08 &  $2.21_{-0.04}^{+0.04}$  &  $4.27_{-0.03}^{+0.03}$ & $0.59_{-0.01}^{+0.01}$  &  $3.97_{-0.19}^{+0.21}$ &   $1.60_{-0.30}^{+0.31}$ &  48.50 \\
		20078-01-04-00  &  50478.08 &  $2.18_{-0.04}^{+0.04}$  &  $4.31_{-0.03}^{+0.03}$ & $0.58_{-0.01}^{+0.01}$  &  $4.17_{-0.19}^{+0.21}$ &   $1.37_{-0.28}^{+0.28}$ &  46.86 \\
		20077-01-09-00  &  50480.82 &  $2.22_{-0.04}^{+0.04}$  &  $4.27_{-0.03}^{+0.03}$ & $0.61_{-0.01}^{+0.01}$  &  $3.87_{-0.20}^{+0.22}$ &   $1.82_{-0.33}^{+0.34}$ &  45.32 \\
		20077-01-10-00  &  50482.61 &  $2.12_{-0.05}^{+0.06}$  &  $4.26_{-0.04}^{+0.04}$ & $0.62_{-0.01}^{+0.01}$  &  $3.95_{-0.33}^{+0.37}$ &   $1.76_{-0.49}^{+0.51}$ &  45.82 \\
		20078-01-05-00  &  50484.62 &  $2.17_{-0.06}^{+0.03}$  &  $4.28_{-0.04}^{+0.03}$ & $0.60_{-0.01}^{+0.01}$  &  $4.10_{-0.24}^{+0.26}$ &   $1.52_{-0.35}^{+0.36}$ &  44.82 \\
		20078-01-05-01  &  50484.75 &  $2.21_{-0.08}^{+0.03}$  &  $4.33_{-0.05}^{+0.03}$ & $0.60_{-0.01}^{+0.01}$  &  $3.83_{-0.24}^{+0.26}$ &   $1.90_{-0.39}^{+0.40}$ &  46.02 \\
		20078-01-05-02  &  50485.09 &  $2.20_{-0.04}^{+0.04}$  &  $4.34_{-0.03}^{+0.03}$ & $0.60_{-0.01}^{+0.01}$  &  $4.08_{-0.21}^{+0.22}$ &   $1.59_{-0.31}^{+0.32}$ &  42.66 \\
		20077-01-11-00  &  50487.89 &  $2.15_{-0.05}^{+0.06}$  &  $4.29_{-0.04}^{+0.04}$ & $0.60_{-0.01}^{+0.01}$  &  $4.06_{-0.29}^{+0.32}$ &   $1.56_{-0.42}^{+0.43}$ &  39.91 \\
		20077-01-12-00  &  50489.49 &  $2.19_{-0.06}^{+0.04}$  &  $4.33_{-0.04}^{+0.03}$ & $0.59_{-0.01}^{+0.01}$  &  $4.06_{-0.23}^{+0.25}$ &   $1.57_{-0.33}^{+0.34}$ &  39.96 \\
		20078-01-06-00  &  50492.10 &  $2.23_{-0.06}^{+0.03}$  &  $4.36_{-0.04}^{+0.02}$ & $0.58_{-0.01}^{+0.01}$  &  $4.05_{-0.19}^{+0.21}$ &   $1.57_{-0.29}^{+0.30}$ &  38.00 \\
		20078-01-06-01  &  50492.23 &  $2.18_{-0.04}^{+0.04}$  &  $4.36_{-0.03}^{+0.03}$ & $0.58_{-0.01}^{+0.01}$  &  $4.15_{-0.21}^{+0.22}$ &   $1.46_{-0.30}^{+0.31}$ &  37.47 \\
		20078-01-06-02  &  50492.43 &  $2.17_{-0.03}^{+0.07}$  &  $4.29_{-0.03}^{+0.04}$ & $0.59_{-0.01}^{+0.01}$  &  $4.09_{-0.23}^{+0.24}$ &   $1.47_{-0.33}^{+0.34}$ &  37.93 \\
		20078-01-07-00  &  50494.73 &  $2.19_{-0.03}^{+0.05}$  &  $4.34_{-0.02}^{+0.03}$ & $0.58_{-0.01}^{+0.01}$  &  $4.18_{-0.18}^{+0.19}$ &   $1.38_{-0.26}^{+0.27}$ &  35.30 \\
		20077-01-13-00  &  50497.24 &  $2.18_{-0.05}^{+0.06}$  &  $4.40_{-0.04}^{+0.03}$ & $0.58_{-0.01}^{+0.01}$  &  $3.93_{-0.25}^{+0.28}$ &   $1.79_{-0.39}^{+0.41}$ &  32.36 \\
		20077-01-14-00  &  50499.97 &  $2.15_{-0.06}^{+0.04}$  &  $4.37_{-0.04}^{+0.02}$ & $0.57_{-0.01}^{+0.01}$  &  $4.11_{-0.29}^{+0.33}$ &   $1.47_{-0.41}^{+0.42}$ &  31.91 \\
		20077-01-15-00  &  50501.83 &  $2.14_{-0.04}^{+0.05}$  &  $4.40_{-0.03}^{+0.03}$ & $0.57_{-0.01}^{+0.01}$  &  $4.28_{-0.24}^{+0.26}$ &   $1.32_{-0.32}^{+0.32}$ &  30.34 \\
		20077-01-16-00  &  50503.83 &  $2.14_{-0.06}^{+0.04}$  &  $4.39_{-0.04}^{+0.03}$ & $0.57_{-0.01}^{+0.01}$  &  $4.32_{-0.24}^{+0.26}$ &   $1.24_{-0.31}^{+0.32}$ &  29.58 \\
		20078-01-08-01  &  50505.11 &  $2.16_{-0.04}^{+0.04}$  &  $4.38_{-0.03}^{+0.03}$ & $0.56_{-0.01}^{+0.01}$  &  $4.35_{-0.20}^{+0.21}$ &   $1.15_{-0.26}^{+0.27}$ &  28.61 \\
		20078-01-08-00  &  50505.23 &  $2.16_{-0.05}^{+0.03}$  &  $4.39_{-0.04}^{+0.02}$ & $0.56_{-0.01}^{+0.01}$  &  $4.33_{-0.21}^{+0.23}$ &   $1.21_{-0.28}^{+0.29}$ &  27.97 \\
		20077-01-17-00  &  50509.04 &  $2.11_{-0.06}^{+0.05}$  &  $4.41_{-0.05}^{+0.03}$ & $0.56_{-0.01}^{+0.01}$  &  $4.46_{-0.29}^{+0.32}$ &   $1.10_{-0.36}^{+0.37}$ &  25.29 \\
		20077-01-18-00  &  50511.04 &  $2.10_{-0.05}^{+0.04}$  &  $4.43_{-0.04}^{+0.03}$ & $0.55_{-0.01}^{+0.01}$  &  $4.56_{-0.25}^{+0.27}$ &   $0.99_{-0.29}^{+0.30}$ &  24.56 \\
		20078-01-09-00  &  50513.79 &  $2.09_{-0.05}^{+0.04}$  &  $4.43_{-0.04}^{+0.03}$ & $0.55_{-0.01}^{+0.01}$  &  $4.65_{-0.24}^{+0.26}$ &   $0.89_{-0.28}^{+0.29}$ &  23.28 \\
		20078-01-09-01  &  50513.90 &  $2.10_{-0.05}^{+0.04}$  &  $4.44_{-0.04}^{+0.03}$ & $0.55_{-0.01}^{+0.01}$  &  $4.57_{-0.25}^{+0.27}$ &   $1.00_{-0.29}^{+0.30}$ &  22.96 \\
		20077-01-19-00  &  50516.97 &  $2.08_{-0.05}^{+0.04}$  &  $4.45_{-0.04}^{+0.03}$ & $0.55_{-0.01}^{+0.01}$  &  $4.70_{-0.25}^{+0.27}$ &   $0.87_{-0.28}^{+0.29}$ &  20.42 \\
		20078-01-10-00  &  50517.84 &  $2.05_{-0.05}^{+0.05}$  &  $4.50_{-0.05}^{+0.04}$ & $0.56_{-0.01}^{+0.01}$  &  $4.72_{-0.27}^{+0.29}$ &   $0.96_{-0.30}^{+0.31}$ &  19.44 \\
		20078-01-10-01  &  50518.11 &  $2.05_{-0.04}^{+0.05}$  &  $4.45_{-0.05}^{+0.04}$ & $0.55_{-0.01}^{+0.01}$  &  $4.77_{-0.26}^{+0.28}$ &   $0.85_{-0.28}^{+0.29}$ &  19.08 \\
		20077-01-20-00  &  50520.98 &  $2.00_{-0.05}^{+0.05}$  &  $4.47_{-0.06}^{+0.05}$ & $0.55_{-0.01}^{+0.01}$  &  $4.92_{-0.33}^{+0.36}$ &   $0.77_{-0.33}^{+0.34}$ &  17.26 \\
		20401-01-01-00  &  50523.70 &  $2.01_{-0.05}^{+0.05}$  &  $4.52_{-0.05}^{+0.05}$ & $0.56_{-0.01}^{+0.01}$  &  $4.85_{-0.31}^{+0.35}$ &   $0.91_{-0.33}^{+0.34}$ &  15.49 \\
		20401-01-02-00  &  50525.84 &  $1.99_{-0.04}^{+0.05}$  &  $4.51_{-0.06}^{+0.05}$ & $0.56_{-0.01}^{+0.01}$  &  $5.05_{-0.33}^{+0.37}$ &   $0.73_{-0.33}^{+0.34}$ &  14.40 \\
		20078-01-11-01  &  50527.64 &  $1.97_{-0.05}^{+0.05}$  &  $4.51_{-0.06}^{+0.06}$ & $0.56_{-0.01}^{+0.01}$  &  $4.98_{-0.37}^{+0.42}$ &   $0.83_{-0.37}^{+0.39}$ &  13.13 \\
		20078-01-11-00  &  50527.77 &  $1.99_{-0.05}^{+0.05}$  &  $4.46_{-0.06}^{+0.05}$ & $0.56_{-0.01}^{+0.01}$  &  $4.88_{-0.34}^{+0.38}$ &   $0.85_{-0.36}^{+0.37}$ &  13.12 \\
		20078-01-11-02  &  50527.84 &  $1.95_{-0.05}^{+0.05}$  &  $4.49_{-0.07}^{+0.07}$ & $0.59_{-0.01}^{+0.01}$  &  $5.07_{-0.45}^{+0.53}$ &   $0.85_{-0.45}^{+0.47}$ &  11.81 \\
		20401-01-03-00  &  50528.78 &  $1.98_{-0.05}^{+0.05}$  &  $4.52_{-0.06}^{+0.06}$ & $0.57_{-0.01}^{+0.01}$  &  $5.09_{-0.38}^{+0.42}$ &   $0.76_{-0.37}^{+0.38}$ &  12.67 \\
		20401-01-04-00  &  50531.92 &  $1.94_{-0.05}^{+0.05}$  &  $4.50_{-0.07}^{+0.07}$ & $0.58_{-0.01}^{+0.01}$  &  $5.13_{-0.44}^{+0.51}$ &   $0.77_{-0.42}^{+0.44}$ &  10.46 \\
		20078-01-12-00  &  50534.82 &  $1.92_{-0.04}^{+0.04}$  &  $4.55_{-0.07}^{+0.07}$ & $0.59_{-0.01}^{+0.01}$  &  $5.48_{-0.46}^{+0.53}$ &   $0.63_{-0.40}^{+0.41}$ &  8.53 \\
		20078-01-12-01  &  50534.99 &  $1.91_{-0.05}^{+0.06}$  &  $4.54_{-0.09}^{+0.08}$ & $0.60_{-0.01}^{+0.01}$  &  $5.29_{-0.58}^{+0.70}$ &   $0.80_{-0.54}^{+0.56}$ &  8.40 \\
		20401-01-05-00  &  50536.51 &  $1.96_{-0.05}^{+0.05}$  &  $4.56_{-0.07}^{+0.06}$ & $0.60_{-0.01}^{+0.01}$  &  $4.85_{-0.42}^{+0.49}$ &   $1.19_{-0.46}^{+0.48}$ &  7.79 \\
		20401-01-06-00  &  50538.38 &  $1.88_{-0.05}^{+0.05}$  &  $4.47_{-0.10}^{+0.09}$ & $0.62_{-0.01}^{+0.01}$  &  $5.40_{-0.68}^{+0.85}$ &   $0.82_{-0.62}^{+0.65}$ &  6.40 \\
		20078-01-13-01  &  50541.47 &  $1.88_{-0.05}^{+0.06}$  &  $4.40_{-0.14}^{+0.13}$ & $0.66_{-0.01}^{+0.01}$  &  $5.54_{-1.00}^{+1.34}$ &   $<1.83$                &  4.95 \\
		20078-01-13-00  &  50541.60 &  $1.91_{-0.04}^{+0.04}$  &  $4.41_{-0.09}^{+0.09}$ & $0.65_{-0.01}^{+0.01}$  &  $5.77_{-0.69}^{+0.85}$ &   $0.62_{-0.57}^{+0.60}$ &  4.85 \\
		20401-01-07-00  &  50543.51 &  $1.88_{-0.04}^{+0.05}$  &  $4.20_{-0.14}^{+0.12}$ & $0.66_{-0.01}^{+0.01}$  &  $6.47_{-1.10}^{+0.28}$ &   $0.91$                 &  4.25 \\
		20401-01-08-00  &  50548.19 &  $1.91_{-0.05}^{+0.06}$  &  $4.17_{-0.23}^{+0.19}$ & $0.74_{-0.01}^{+0.01}$  &  $6.61_{-1.83}^{+0.84}$ &   $<1.89$                &  2.89 \\
		20078-01-14-01  &  50549.06 &  $1.97_{-0.06}^{+0.07}$  &  $4.21_{-0.18}^{+0.16}$ & $0.76_{-0.02}^{+0.01}$  &  $4.76_{-1.21}^{+1.92}$ &   $1.79_{-1.54}^{+1.79}$ &  2.76 \\
		20078-01-14-00  &  50549.44 &  $1.89_{-0.04}^{+0.04}$  &  $4.01_{-0.24}^{+0.20}$ & $0.75_{-0.00}^{+0.01}$  &  $7.50_{-1.86}^{+0.28}$ &   $<1.20$                &  2.66 \\
		20401-01-09-00  &  50553.12 &  $1.94_{-0.05}^{+0.05}$  &  $3.72_{-0.25}^{+0.21}$ & $0.78_{-0.01}^{+0.01}$  &  $6.94_{-1.65}^{+0.31}$ &   $<1.17$                &  2.16 \\
		20401-01-10-00  &  50560.20 &  $2.00_{-0.04}^{+0.05}$  &  $3.18_{-0.22}^{+0.20}$ & $0.82_{-0.01}^{+0.01}$  &  $6.19_{-0.62}^{+0.25}$ &   $<0.43$                &  1.63 \\
		20078-01-16-01  &  50561.93 &  $2.17_{-0.15}^{+0.21}$  &  $3.52_{-0.55}^{+0.37}$ & $0.88_{-0.04}^{+0.03}$  &  $5.22_{-3.15}^{+2.31}$ &   $<7.94$                &  1.48 \\
		20078-01-16-00  &  50562.15 &  $2.00_{-0.04}^{+0.04}$  &  $3.01_{-0.23}^{+0.22}$ & $0.84_{-0.01}^{+0.01}$  &  $6.23_{-0.49}^{+0.27}$ &   $<0.34$                &  1.50 \\
		20078-01-16-02  &  50562.88 &  $1.94_{-0.05}^{+0.07}$  &  $2.71_{-0.40}^{+0.44}$ & $0.84_{-0.02}^{+0.02}$  &  $6.40_{-1.08}^{+0.51}$ &   $<0.78$                &  1.51 \\
		20401-01-12-00  &  50565.87 &  $2.05_{-0.05}^{+0.06}$  &  $2.70_{-0.33}^{+0.34}$ & $0.90_{-0.02}^{+0.02}$  &  $6.13_{-1.07}^{+0.46}$ &   $<0.87$                &  1.27 \\
		\hline
		\caption{Spectral fits of GRO J1744-28 by using "wabs*(gauss+powerlaw+bbodyrad)" and "wabs*(gauss+comptb)" models, respectively.}
		\label{longtable}
	\end{longtable}
}

\end{document}